\documentclass[twocolumn,showpacs,preprintnumbers,amsmath,amssymb,prb,superscriptaddress]{revtex4}
\usepackage{graphicx}
\usepackage{amsmath}
\usepackage{bm}
\usepackage{longtable}

\begin{document}

\title{Optical polarization of localized hole spins in $p$-doped quantum wells}

\author{M. Studer}
\affiliation{Solid State Physics Laboratory, ETH Zurich, 8093
Zurich, Switzerland} \affiliation{IBM Research--Zurich,
S\"aumerstrasse 4, 8803 R\"uschlikon, Switzerland}
\author{M. Hirmer}
\affiliation{Institut f\"ur Experimentelle und Angewandte Physik,
Universit\"at Regensburg, 93040 Regensburg, Germany}
\author{D. Schuh}
\affiliation{Institut f\"ur Experimentelle und Angewandte Physik,
Universit\"at Regensburg, 93040 Regensburg, Germany}
\author{W. Wegscheider}
\affiliation{Solid State Physics Laboratory, ETH Zurich, 8093
Zurich, Switzerland}
\author{K. Ensslin}
\affiliation{Solid State Physics Laboratory, ETH Zurich, 8093
Zurich, Switzerland}
\author{G. Salis\footnote{gsa@zurich.ibm.com}}
\affiliation{IBM Research--Zurich, S\"aumerstrasse 4, 8803
R\"uschlikon, Switzerland}
\begin{abstract}

The initialization of spin polarization in localized hole states is
investigated using time-resolved Kerr rotation. We find that the sign
of the polarization depends on the magnetic field, and the power and
the wavelength of the circularly polarized pump pulse. An analysis
of the spin dynamics and the spin-initialization process shows that two mechanisms are responsible for spin polarization with opposite sign: The difference of the $g$ factor between the localized holes and the
trions, as well as the capturing process of dark excitons by the
localized hole states.
\end{abstract}

\maketitle

Localized hole spins in $p$-doped III-V semiconductors can have
considerably long spin lifetimes\cite{Heiss2007} in the range of
100\,$\mu$s and coherence times\cite{Brunner2009} on the order of
$\mu$s. Holes or electrons can be localized in natural quantum dots
(QDs) formed due to potential fluctuations in quantum wells
(QWs).\cite{Zrenner1994,Hess1994} The spin of localized hole states
in p-doped GaAs/AlGaAs QWs has been studied using time-resolved Kerr
rotation (TRKR).\cite{Syperek2007,Kugler2009,Korn2009} Information
on hole spins can also be obtained in n-doped QW structures by
studying the recombination of an optically excited hole spin with a
resident electron.\cite{Marie1999,Yugova2009} The reliable
polarization of hole spins in QDs is one of the key requirements
necessary to study this quantum system. Possible polarization
mechanisms make use of the properties of the optically excited
positively charged trion that are different compared to those of the
bare hole, e.g. the different interaction with nuclear
spins\cite{Gerardot2008} or the different $g$ factor.\cite{Korn2009}

In our study, a $p$-doped QW is excited with circularly polarized
photons giving rise to two competing spin polarization mechanisms
leading to polarization of an ensemble of localized holes. The two
mechanisms polarize the spins with opposite signs, and their
relative strength depends on the external magnetic field
($B_\textrm{ext}$), the pump power ($P_\textrm{P}$) and the
wavelength ($\lambda $) of the pump beam. The first mechanism is
found to rely on the difference of the trion and the hole $g$
factors and disappears for $B_\textrm{ext}=0$. The second mechanism
remains effective for $B_\textrm{ext}=0$ and strongly depends on
$P_\textrm{P}$ and $\lambda$. The relative strength of the two
mechanisms can therefore be controlled by the properties of the pump
pulse. We show that the second mechanism can be
explained by a capturing of dark excitons by the localized holes.

\begin{figure}[htb]
\includegraphics[width=84mm]{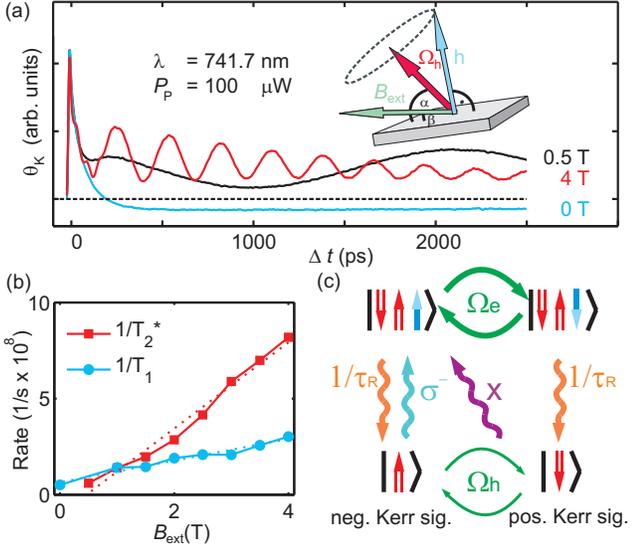}
\caption{\label{fig:fig1}(Color online) (a) Experimental TRKR
signals for three $B_\textrm{ext}$ values for a probe power of
20~$\mu$W. The insert shows a schematic precession of the hole spins
(h) in the tilted $B_\textrm{ext}$ around the precession axis
$\Omega_h$. (b) Dephasing rates vs. $B_\textrm{ext}$. (c) Localized
hole and trion states: Arrows symbolize the transitions between the
four states.}
\end{figure}

To generate and study the spin polarization, we employ
time-resolved Kerr rotation (TRKR). Circularly polarized laser
pulses focused onto a spot with a diameter of about 40~$\mu$m pump
the optical transitions and generate spin-polarized excitons and trions. The generation of a trion with an electron spin up using a $\sigma^-$-polarized photon is depicted in Fig~\ref{fig:fig1}(c).
After the decay of these excited states, an ensemble of localized
hole spins remains polarized (see
discussion below). The polarization of these
hole spins is detected using linearly
polarized probe pulses that are delayed by a time $\Delta t$ with respect to the pump pulses. The reflected probe pulses are analyzed and
reveal the component of spin polarization along the QW growth direction $\mathbf z$ at time $\Delta t$ by a small rotation $\Theta_{\textrm K}$ of the polarization
angle. We use a cascaded lock-in technique where both the circular
polarization of the pump pulse as well as the probe intensity are
modulated. The sample is mounted in a cryostat and cooled to 1.6~K.
The external magnetic field is applied at an angle $\beta=3^\circ$
with respect to the sample plane [see insert of
Fig.~\ref{fig:fig1}(a)]. We investigate a 4-nm-wide remotely doped
GaAs/AlGaAs QW with a hole sheet density of $1.1 \times
10^{15}$~m$^{-2}$ and a mobility of 1.3 m$^2$(Vs)$^{-1}$ measured at
1.3~K.

Figure~\ref{fig:fig1}(a) shows experimental TRKR signals at three
different magnetic fields. At $B_\textrm{ext}>0$, $\Theta_{\textrm
K}$ is the sum of two exponentially decaying cosine functions and a
non-oscillating exponential function. The short-lived oscillation
(best seen at $B_\textrm{ext}=4$~T and $0<\Delta t<100$~ps) is
attributed~\cite{Syperek2007} to the trion spin which is determined
by the spin of the electron in the trion and therefore precesses
with the electron $g$ factor $g_e$. We measure a decay time of
$\tau_\textrm{T}$=80~ps and $g_e$=0.34. We assume that this part of
the signal decays mainly due to the recombination rate of the trions
($1/\tau_R$)~\cite{Kugler2009}, i.e.
$\tau_\textrm{T}\cong\tau_\textrm{R}$. The longer-lived part of
$\Theta_{\textrm K}$ originates from the localized hole
spins.~\cite{Syperek2007} Due to the tilt angle $\beta$ and the
strong anisotropy of the hole $g$ factor, the precession axis of the
hole spins is tilted out of plane\cite{Korn2009} by an angle
$\alpha>\beta$ [see insert of Fig.~\ref{fig:fig1}(a)]. The Kerr
signal is proportional to the projection of the spins of the
ensemble along $\mathbf z$ and therefore also has a non-oscillating
part. Note that for the two curves at $B_\textrm{ext}=0.5$ and 4\,T,
the trion and both parts of the hole signal have a positive
amplitude.

The situation for $B_\textrm{ext}$=0 is strikingly different. There,
$\Theta_{\textrm K}$ can be described by a superposition of two exponentials
with amplitudes of opposite sign. While the short-lived trion signal is still positive, the long-lived signal from the localized holes now has a negative
amplitude. From this, we conclude that there must be a
$B_\textrm{ext}$-dependent sign change of the initialized hole
polarization. We have compared Kerr signals for finite and zero
magnetic fields for various $\lambda$ and QWs with widths between 4
and 15~nm and always found a sign-reversal of these hole spins.

We now proceed to analyze the observed dynamics in more detail. The
TRKR signal of the localized holes (neglecting the fast-decaying component from the trion spin) is described by
\begin{equation}
\label{Eq:HoleSig} \Theta_\textrm{K}=  A \cdot I_h \big[a_1  e^{-\Delta t/T_1} + a_2
e^{-\Delta t/T_2^*} \cos(\Omega_h \Delta t)\big],
\end{equation}
where $\Omega_h=g_h \mu_B B_\textrm{ext}$. Here, $A$ is the
$\lambda$-dependent amplitude of the Kerr rotation and
$g_h\approx0.06$ is the hole $g$ factor. The non-oscillating and oscillating parts are proportional to the projection of the respective spin component onto $\mathbf z$ and are given by $a_1=\sin^2\alpha$ and $a_2=\cos^2\alpha$. They decay with time constants $T_1$ and $T_2^*$. The function $I_\textrm{h}$
describes the effectiveness of the pump pulse to initialize hole
spin polarization. It depends on $B_\textrm{ext}$,
$\lambda$ and $P_\textrm{P}$. The sign change of the hole spin
polarization at low $B_\textrm{ext}$ is attributed to a sign change
of $I_\textrm{h}$.

Figure~\ref{fig:fig1}(b) shows the dephasing rates as a function of
$B_\textrm{ext}$. The decay rates were deduced from fitting
Eq.~\ref{Eq:HoleSig} to $\Theta_\textrm{K}$ in a time window
$100<\Delta t<2500$~ps. After a flat region for small $B_\textrm{ext}$, $1/T_2^*$ increases about linearly with $B_\textrm{ext}$,  which is typical for an inhomogeneous broadening of
the $g$ factor~\cite{Crooker2010}. A fit of the data in Fig.~\ref{fig:fig1}(b) yields a slope of $k_2=2.2 \times 10^8$~s$^{-1}$T$^{-1}$. Also $1/T_1$ increases with $B_\textrm{ext}$ with a slope of $k_1=5.7 \times 10^7$~s$^{-1}$T$^{-1}$.

\begin{figure}[htb]
\includegraphics[width=84mm]{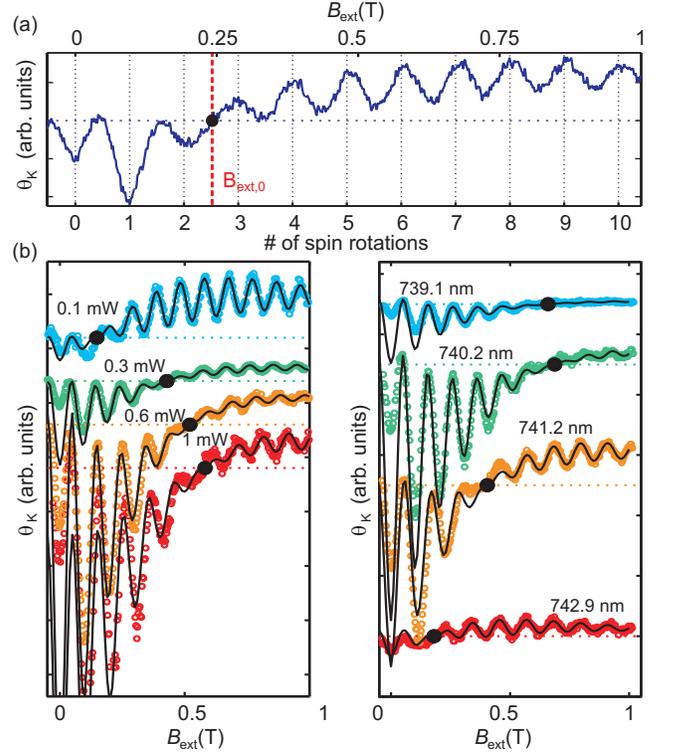}
\caption{\label{fig:fig2}(Color online) Kerr rotation measurements
vs. $B_\textrm{ext}$ at $\Delta t= 12.4$~ns using a probe power of
10~$\mu$W. (a) $\lambda=741.2$~nm and $P_\textrm{P}$=0.2~mW. The
vertical line and the dot mark the $B_\textrm{ext}$ value where
$I_\textrm{h}=0$. (b) left panel: $\lambda=741.2$~nm and varying
$P_\textrm{P}$. Right panel: $P_\textrm{P}=500$~$\mu$W and varying
$\lambda$. For all measurements in (b) the fits are superimposed as
solid lines.}
\end{figure}

To study the initialization at low $B_\textrm{ext}$ and the sign reversal, $\Theta_{\textrm K}$ was measured at fixed time delay $\Delta t=12.4$\,ns as a function of $B_\textrm{ext}$ [see Fig.~\ref{fig:fig2}(a)]. The
measured curves are well described by Eq.~\ref{Eq:HoleSig}, i.e. a
$B_\textrm{ext}$-dependent cosine function, offset by a
non-oscillating part. Note that if $T_2^*$ becomes longer than the laser pulse repetition time of 12.5\,ns, the spin polarization created by the previous pump
pulses becomes important~\cite{Kikkawa1998} and the oscillations in $B_\textrm{ext}$ deviate from a cosine shape. This is observed in our samples only at $B_\textrm{ext}<0.2$\,T, in agreement with the data shown in Fig.~\ref{fig:fig1}(b). The sign change of $I_h$ is very well
resolved: In Fig.~\ref{fig:fig2}(a) we measure dips for
$|B_\textrm{ext}|<0.2$~T at every integer spin rotation whereas for
$|B_\textrm{ext}|>0.3$~T we measure peaks. Between these two regimes
there is a magnetic field $B_\textrm{ext,0}$ where $\Theta_{\textrm K}=0$ (marked with a dot and a dashed line). At this field, the
initialization of hole polarization is ineffective, i.e.
$I_\textrm{h}=0$.

Figure~\ref{fig:fig2}(b) shows Kerr rotation data vs.
$B_\textrm{ext}$ for different $P_\textrm{P}$ (left panel) and
$\lambda$  (right panel). The curves are offset for clarity and the
dashed line indicates $\Theta_\textrm{K}=0$ for every curve. The
black dots mark $B_\textrm{ext,0}$. We find a trend towards higher
values of $B_\textrm{ext,0}$ for increased $P_\textrm{P}$ or lower
$\lambda$. In order to understand and fit the measured curves we now
proceed to calculate $I_\textrm{h}$.

The dynamics of the initialization process involves the transfer
between the four states depicted in Fig.~\ref{fig:fig1}(c). Before
the arrival of a pump pulse, a localized hole spin has the same
probability to be in an up or in a down state $p_{\Delta
t=0}(\mid\Uparrow \rangle)=p_{\Delta t=0}(\mid \Downarrow \rangle
)=0.5$. Without loss of generality, we assume a $\sigma^-$ pulse to
arrive at $\Delta t$=0. Due to optical selection rules, the pulse
pumps $ \mid \Uparrow \rangle $-holes into $\mid \Downarrow \Uparrow
\uparrow \rangle$-trions with a probability $p_\sigma$. We assume
pure heavy-hole states and neglect any interaction of the pump with
light holes. This is justified since light holes are sufficiently
far away in energy. For $B_\textrm{ext}$=0, the trions decay with
the carrier decay rate $1/\tau_\textrm{R}$ into the original spin
state with no resulting polarization. Since $\tau_\textrm{R}\ll
T_2^*, T_1$, the hole spin decay is neglected for calculating
$I_\textrm{h}$. For $B_\textrm{ext}\neq0$, the trion spin precesses
with a frequency $\Omega_e$, whereas the hole spin precesses with a
lower frequency ($\Omega_h$) due to the smaller $g$ factor. The
trions therefore do not decay into the original spin state. The
result is a positive $I_h$ increasing with
$B_\textrm{ext}$.\cite{Korn2009}

In the experiment, we observe a $\mid \Uparrow \rangle$ polarization
for $B_\textrm{ext}=0$, i.e. a negative $I_h$. In order to explain
this observation, we introduce a process $X$ that pumps $\mid
\Downarrow \rangle$-holes into $\mid\Downarrow \Uparrow  \uparrow
\rangle$-trions. We assume this transition to be fast compared to
$\tau_R$. A possible mechanism is discussed below. Both processes
($\sigma$ and $X$) pump localized holes into the $\mid \Downarrow
\Uparrow \uparrow \rangle$-trion state [see Fig.~\ref{fig:fig1}(c)]
and the total occupation probability of this state is given by
\mbox{$p_\sigma p_{\Delta t=0}(\mid\Uparrow \rangle)+ p_X p_{\Delta
t=0}(\mid \Downarrow \rangle )$=}\mbox{$0.5[p_\sigma+p_X]$}. For
$B_\textrm{ext}=0$ and finite $p_X$, this results in a negative Kerr
signal after the decay of the trions.

The system shown in Fig.~\ref{fig:fig1}(c) is described by an
analytically solvable system of rate equations that include the
precession of the spins, the dephasing of the holes and the decay of
the trions.~\cite{Korn2009} Eq.~\ref{Eq:HoleSig} is the solution of
this system and
\begin{equation}
\label{Eq:Init}
 I_\textrm{h}=\frac{ p_\sigma}{2} \Big[1 - \frac{p_X}{p_\sigma} - \frac{ 1
+\frac{p_X}{p_\sigma}}{\Omega_{e-h}^2\tau_R^2+1} \Big].
\end{equation}

In Eq.~\ref{Eq:Init},  $\Omega_{e-h}=|g_e-g_h|\mu_B
B_\textrm{ext}/\hbar$. We assume $p_\sigma$ and $p_X$ to be
independent of $B_\textrm{ext}$. For $B_\textrm{ext}=0$,
we recover $I_\textrm{h}=-p_X$, i.e. the spin-polarization is only determined by the $X$ process. For finite $p_X$, $I_h$ has a zero
point at finite magnetic field ($B_\textrm{ext,0})$. From this
point, the ratio of the two pump probabilities can be calculated
\begin{equation}
\label{Eq:PumpRatio}
\frac{p_{X}}{p_\sigma}=\frac{\Omega_\textrm{e-h,0}^2\tau_\textrm{R}^2}{\Omega_\textrm{e-h,0}^2\tau_\textrm{R}^2+2},
\end{equation}
with $\Omega_\textrm{e-h,0}=|g_e-g_h| \mu_B B_\textrm{ext,0}/\hbar$.

For the next step, we use Eqs.~\ref{Eq:HoleSig}-\ref{Eq:PumpRatio}
to fit the curves in Fig.~\ref{fig:fig2}(b). The spin lifetimes are
modeled by $1/T_1=1/T_{1,0}+k_1 B_\textrm{ext}$ and
$1/T_2^*=1/T^*_{2,0}+k_2 B_\textrm{ext}$. The measured parameters
are $g_\textrm{e}=0.34$, $\tau_\textrm{R}=80$\,ps and
$B_\textrm{ext,0}$ that varies with $P_\textrm{P}$ and determines
through Eq.~\ref{Eq:PumpRatio} the ratio $p_\sigma/p_X$. For each
value of $P_\textrm{P}$, fit parameters $k_1$, $k_2$, $g_h$, $c_1$
and $c_2$ are determined, where $c_1=A p_\sigma a_1 e^{-\Delta
t/T_{1,0}}$ and $c_2=A p_\sigma a_2 e^{-\Delta t/T^*_{2,0}}$ are the
$B_\textrm{ext}$-independent amplitudes of the non-oscillating and
oscillating parts of the signal. The assumed linear increase of the
spin relaxation rate with $B_\textrm{ext}$ overestimates the Kerr
signal below 0.2\,T. To account for this we weighted the
least-square residuals with $B_\textrm{ext}^2$ for fitting the
curves in the range between 0 and 1\,T. The solid black lines in
Fig.~\ref{fig:fig2}(b) are the resulting fits. The agreement between
the data and the fit is very good for $B_\textrm{ext}>0.2$~T.

\begin{figure}[htb]
\includegraphics[width=84mm]{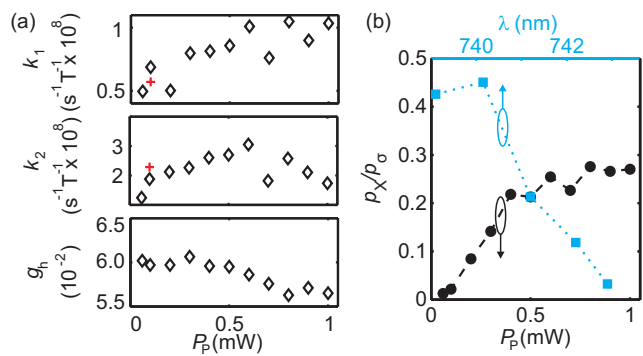}
\caption{\label{fig:fig3} (Color online) (a) Parameters of the fits
vs. pump power $P_\textrm{P}$ as described in the text (diamonds) and for
comparison fit results obtained from the data shown in
Fig.~\ref{fig:fig1}(b) (crosses) (b) Ratio of the two pumping
mechanisms as a function of $P_\textrm{P}$ (dots) and $\lambda$
(squares).}
\end{figure}

The fit parameters as a function of $P_\textrm{P}$ are displayed in
Fig.~\ref{fig:fig3}(a). The values for $k_1$ and $k_2$ match the results shown as a cross that were obtained from measurements of $\Theta_{\textrm K}$ versus $\Delta t$ [see Fig.~\ref{fig:fig1}(a) and (b)]. This demonstrates that the field-dependence of $\Theta_{\textrm K}$ measured in Fig.~\ref{fig:fig2} is well described by our model and yields the same $B_\textrm{ext}$-dependence for both $T_1$ and $T^*_2$. Only small changes on the fit parameters are observed as a function of $P_P$,
reflecting the notion that the pump-power dependence of $I_h$ is the
main cause for the difference between the curves in
Fig.~\ref{fig:fig2}. For increased $P_P$, $g_h$ decreases by almost
10\% which could be attributed to the high sensitivity of $g_h$ on
changes in the electrostatic confinement.\cite{Kugler2009}

From the values of $B_\textrm{ext,0}$, we derive the ratio
$p_X/p_\sigma$. The points plotted in Fig.~\ref{fig:fig3}(b) are an
average of $|B_\textrm{ext,0}|$ for the two points obtained by a
magnetic field sweep from -1~T to 1~T. The black dots show the ratio
as a function of $P_\textrm{P}$ for $\lambda=741.2$~nm. The ratio
increases and saturates at a value of 0.25 for
$P_\textrm{P}>0.5$~mW. By changing the wavelength and keeping the
pump power at 0.5~mW, even larger ratios can be obtained. For
$\lambda$ outside the range of the data shown in
Fig.~\ref{fig:fig3}(b), no hole signal was detected, which we relate
to a decrease of the Kerr sensitivity $A$. The obtained
$p_X/p_\sigma$-ratios depend on $\tau_\textrm{R}$
and would be higher if $\tau_\textrm{T}$ were not limited by the
recombination.

Finally, we discuss the mechanism $X$ that pumps the $\mid \Downarrow
\rangle$-holes to the $\mid \Downarrow \Uparrow \uparrow
\rangle$-trions. The $\sigma^-$-pulses also create free $\mid
\Downarrow \uparrow \rangle$-excitons in the QW. The hole spins in the
excitons have a much higher relaxation rate than the
electron spins~\cite{Damen1991,Baylac1995}, leading to a conversion of
bright excitons into $\mid \Uparrow \uparrow \rangle$-excitons.
These dark excitons can combine with localized $\mid \Downarrow \rangle$-holes~\cite{Shabaev2009,Eble2008} to form a $\mid\Downarrow \Uparrow \uparrow \rangle$-trion. This process results in
a transition $X$,  explaining the negative $I_\textrm{h}$ at low
$B_\textrm{ext}$. Note that the bright excitons can also be captured
by a $\mid \Uparrow \rangle$-hole. In our experiment, this process
can not be differentiated from direct excitation of a $\mid
\Downarrow \Uparrow \uparrow \rangle$-trion and is therefore
included in $p_{\sigma}$.

With these processes we can qualitatively explain the data in
Fig.~\ref{fig:fig3}(b). The trion formation with probability $p_\sigma$ is a process that involves first the generation of an electron-hole pair and then the capturing of a resident hole. Because of the limited availability of resident holes, an increase in $P_\textrm{P}$ decreases the average probability $p_\sigma$ that the created electron-hole pairs can capture $\mid \Uparrow \rangle$ holes, and increases the probability $p_X$ for a spin flip of the photo-created holes, i.e. the formation of dark excitons, and the subsequent capturing of still available $\mid \Downarrow \rangle$ hole spins. This explains the increase
of $p_X/p_\sigma$ with pump power. The saturation at
$P_\textrm{P}=0.5$~mW is governed by the rates describing the
capturing and the spin-flip processes of the excitons. At this pump power, the finite number of localized holes also limits $p_X$ for very
large $P_\textrm{P}$. A shorter $\lambda$ increases the
$p_X/p_\sigma$-ratio for the same reason: The absorption of photons increases for shorter $\lambda$, and more electron-hole pairs are generated at same pump power. In addtion, a more pronounced hole dephasing for non-resonantly
pumped excitons~\cite{Damen1991} may favor process $X$ with decreasing $\lambda$. The oscillation in the data of $\Theta_\textrm{K}$ at zero field in Fig.~\ref{fig:fig2}(b) has a
smaller amplitude than the fits, showing the limitation of
our model that assumes $p_X$ to be independent of $B_\textrm{ext}$ and neglects the time-evolution of the excitonic states. The smaller peak is compatible with assuming that the formation of
dark excitons is suppressed for small $B_\textrm{ext}$. A reason for this could be the spin dynamics of the excitons under the combined influence of both an external and anisotropic exchange splitting~\cite{Blackwood1994,Gammon1996}.

To conclude, we find that localized hole spins can be spin-polarized
by exploiting the difference in  $g$ factor between holes and trions
or via the capturing of dark excitons. These two mechanisms lead to the initialization of spin polarization of opposite signs, and their relative strength can be controlled by magnetic field, pump power and wavelength of the pump pulses. By changing these
parameters, the size and sign of the spin polarization of localized holes can be controlled.

We thank the SNF and the KTI for financial support. Stimulating discussion with R. Allenspach, A. Fuhrer, T. Korn, M. Kugler, M. Syperek and D. R. Yakovlev are acknowledged.

\bibliographystyle{prsty}

\end{document}